\documentclass[aps,twocolumn,lengthcheck,preprintnumbers,superscriptaddress]{revtex4-1}%
\usepackage{amsfonts}
\usepackage{amsmath}
\usepackage{amssymb}
\usepackage{graphicx}
\usepackage{color}
\usepackage[mathcal]{euscript}

\usepackage{txfonts}
\usepackage[unicode]{hyperref}
\hypersetup{
   unicode=true,          % non-Latin characters in AcrobatÕs bookmarks
   plainpages=false,
   colorlinks=true,       % false: boxed links; true: colored links
   citecolor=blue,        % color of links to bibliography
}

\begin{document}

\author{J. A. Gil Granados}
\affiliation{Departament de F\'isica Qu\`antica i Astrof\'isica, 
Facultat de F\'{\i}sica, Universitat de Barcelona, E--08028
Barcelona, Spain}
\affiliation{Institut de Ci\`encies del Cosmos, Universitat de Barcelona, 
ICCUB, 08028-Barcelona, Spain}
\author{A. Mu\~{n}oz Mateo}
\affiliation{Departamento de F\'{i}sica, Universidad de La Laguna, E--38200 La Laguna, Spain}
\affiliation{Instituto Universitario de Estudios Avanzados, IUDEA, Universidad de La Laguna, E--38205 La Laguna, Spain}
\author{X. Vi\~nas}
\affiliation{Departament de F\'isica Qu\`antica i Astrof\'isica, 
Facultat de F\'{\i}sica, Universitat de Barcelona, E--08028
Barcelona, Spain}

\title{Half-vortex states in the rotating outer core of neutron stars}

\begin{abstract} 
We probe the superfluid-superconductor dynamics of the rotating outer core of neutron stars through 
half-vortex states. By means of a generalized hydrodynamic model, where proton and neutron fluids are coupled by both dynamic entrainment and Skyrme SLy4 nucleon-nucleon interactions, we analyze single flux tubes in the proton-superconductor component of the system that thread proton vortices located faraway from neutron vortices. It is shown how they give rise to hydrodynamic perturbations in the coexisting neutron superfluid, and its structure remains unaltered for varying rotation rates and magnetic fields within ranges of observational values.
\end{abstract}

\maketitle

\section{Introduction}

Quantized vortices are the signature of superfluidity in quantum degenerated matter systems. They were first observed in type II 
superconductors in the presence of a magnetic field \cite{Essmann1967}, and in superfluid Helium when subjected to rotation \cite{Yarmchuk1979}.  Much later, in 
ultracold atomic gases under highly controlled experimental conditions, isolated quantum vortices, either singly \cite{Matthews1999,Rosenbusch2002} or multiply 
charged \cite{Leanhardt2002}, and vortex arrays \cite{Madison2000,Chevy2000,Abo2001} were observed in Bose-Einstein condensates (BECs), and also
in strongly interacting quantum degenerate Fermi gases \cite{Zwierlein2005,Schunck2007}, even in the presence of spin-population imbalance  
\cite{Zwierlein2006}. In these systems, isolated vortices can be generated by phase imprinting techniques, while vortex arrays were usually produced by rotating 
the atomic cloud. More recently, the realization of synthetic gauge potentials, which simulate electromagnetic fields acting on neutral atoms, opened a 
new path to observe quantum vortex arrays \cite{Lin2009}.

Both agents of vortex excitation, rotation and magnetic field, as high as $10^2-10^3$ Hz and $10^{12}-10^{15}$ G, respectively \cite{Mendell1998}, are naturally present in neutron stars \cite{Baym1969}, whose inner
 layers, crust and core, are hypothesized to show superfluid dynamics. This assumption was made plausible after the observation of the rotational frequency
of pulsars, as measured from radio wave signals \cite{Lyne2000}. It showed sudden spin-ups followed by relatively large relaxation times, of the order of
days or months. In order to explain these events, called pulsar \textit{glitches}, Anderson and Itoh suggested almost fifty years ago 
\cite{Anderson1975} that they could be due to vortex creeping in the inner crust of neutron stars. This layer consists of positively charged 
nuclear clusters, with different geometries, embedded in free electron and neutron fluids at temperatures of the order of $10^8$ K; neutrons achieve sub-nuclear saturation densities, and they are expected to be in a superfluid state \cite{Baym1969}. 

Deeper below the crust, the star's outer core 
involves at least three intermingled fluids: superfluid neutrons, superconductor protons, and a normal, but relativistic, electron fluid. The neutron and proton 
components are mutually coupled, apart from nucleon-nucleon interactions, by entrainment effects, as it happens in superfluid mixtures of $^4$He 
and $^3$He \cite{Andreev1976}. Also as in superfluid Helium or in ultracold gases, the natural rotation of the neutron superfluid triggers the formation  of an array of quantized vortices beyond a rotation threshold. In addition, as in type II superconductors, strong magnetic fields are believed to have induced an array of metastable quantized flux tubes 
(or Abrikosov vortices) in the superconducting protons since early stages of the neutron star formation. 

\begin{figure}[tb]
	\includegraphics[width=1\linewidth]{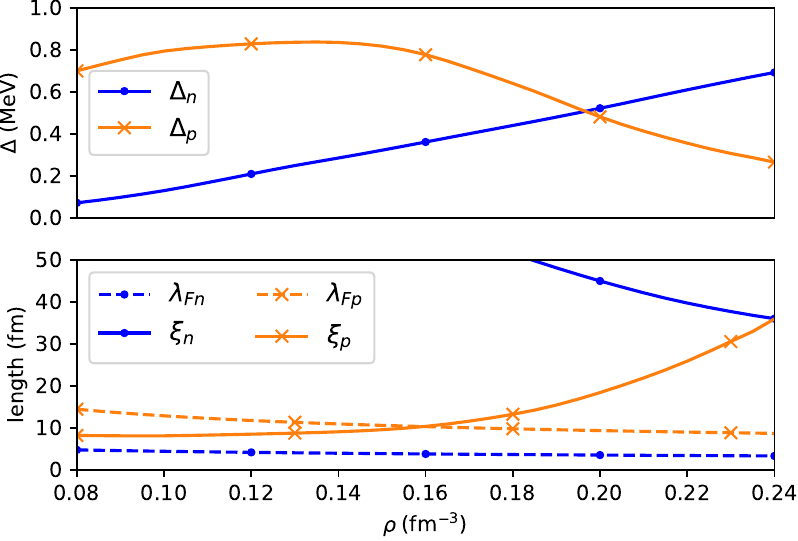}
	\caption{Fermionic pairing and characteristic length scales in the outer core. The Fermi wavelengths $\lambda_{F_\sigma}$ and the coherence lengths $\xi_\sigma$, with $\sigma=p,\,n$, are estimated from the superfluid densities $\rho_\sigma$ and the fermionic pairings $\Delta_\sigma$ (top panel, as reported in Ref. \cite{Zhou2004}).}
	\label{fig:lengths}
\end{figure}
As we know from the BCS theory \cite{Bardeen1957}, superconductivity  in Fermi fluids (or superfluidity for neutral particles) is driven by an 
emerging bosonic pairing field, the order parameter, associated with the creation of Cooper pairs of fermions, and mediated by interparticle interactions, which causes an energy gap in the spectrum of excitations. 
For electrically charged systems, the electromagnetic gauge field combines with the $U(1)$ symmetry of the order parameter to preclude collective excitations in the energy gap \cite{Anderson2015}. In neutron stars, the strong nuclear interaction provides several channels in which such a pairing is possible.
 At lower densities, in the inner crust, the Cooper pairs of neutrons should be mainly in the $^1S_0$ state \cite{Sauls1989}. In the outer core, 
where neutrons and protons settle in $\beta -$equilibrium, superfluid protons remain also at "small" densities and so again they should pair in the 
$^1S_0$ channel. However, since outer-core-neutron densities are about or larger than nuclear saturation density, the neutron superfluid would 
mainly occur in the $^3P_2-^3F_2$ channel.  Maximum pairing gaps reached by $S$-proton pairs, as computed with microscopic interactions, are about $\Delta_p\sim 1$ MeV at baryon densities  $\rho_b\sim 0.12$ fm$^{-3}$, which corresponds to a Fermi wave
 number $k_{Fp} \sim 0.7-0.8$ fm$^{-1}$,  while for the $PF$-neutron pairs the numerical results produce clearly smaller values 
$\Delta_n\sim 0.6$ MeV at $\rho_b\sim 0.24$ fm$^{-3}$, with $k_{Fn} \sim$ 2 fm $^{-1}$ \cite{Baldo1998,Zhou2004} (see Fig. \ref{fig:lengths}).

Pairing fields $\Delta_\sigma$ and particle densities $\rho_\sigma$ (where $\sigma=n,\,p$, correspond to neutrons and protons, respectively) are associated with characteristic length scales: coherence (or healing) length $\xi_\sigma=\hbar^2 k_F/(\pi m \Delta_\sigma)$ and 
Fermi wavelength $\lambda_{F\sigma}=2\pi k_{F\sigma}^{-1}$, respectively, that shape the vortex lines of the superfluids. The typical vortex core size  (see, e.g., 
Ref. \cite{DeGennes}) is of the order of the coherence length $\xi_\sigma$, whereas density inhomogeneities are 
characterized by the bulk value of the Fermi wavelength $\lambda_F$ ( see Fig. \ref{fig:lengths} for the typical values in the outer core).
Numerical calculations in the BCS-BEC crossover of ultracold atoms show, by solving
 self-consistently the Bogoliubov-de Gennes equations \cite{Sensarma2006,Simonucci2013}, how the interplay of these lengths is particularly 
manifest in the structure of the vortex core within the BCS limit. In this latter regime, although the pairing vanishes at the vortex core, the 
fermionic density does not, due to the localization of fermionic Andreev (also named Caroli–de Gennes-Matricon) bound states \cite{Andreev1966,
DeGennes}. This is also the situation found in neutron stars, as shown, for instance, in Ref.~\cite{Yu2003}, where the spatial structure of a vortex is obtained by a self-consistent approach at low neutron-matter density in the inner crust. It was found that the profile of the pairing field, related to  the order parameter, vanishes at the vortex axis, while the matter density only shows a partial depletion \cite{Pecak2021}. 
Notice that this fermion-vortex features contrast
with those of vortices in bosonic diluted gases, where both, pairing field and mat-
ter density profiles, vanish at the vortex axis owing to the fact
that the order parameter is directly related to the bosonic density.
The aforementioned fermion-vortex features have been also reported in other microscopic calculations of isolated vortices in the inner crust \cite{Blasio1999,Avogadro2007}, including finite temperature effects \cite{Pecak2021} and vortex dynamics \cite{Wlazlowski2016}. However, as far as we know, the study of vortex structures in the outer core, where the superfluid dynamics involves a 
two-component system, are at least scarce, and mainly focus on the collective effect of vortices and vortex lattices 
 \cite{Mendell1991,Mendell1991b,Mendell1998,Andersson2005,Glampedakis2011,Link2012,Sourie2020,Sourie2020b,Wood2022}, though some features related to the energy and the magnetic field of single flux tubes in rotating  superconductors have been revealed \cite{Carter2000,Prix2000,Glampedakis2011}.

 Since the rotation rate and the magnetism of a neutron star are assumed to produce vortex lattices in its superfluid layers, additional relevant length scales arise from the separation between neutron and proton vortices in these lattices, which are estimated to be \cite{Glampedakis2011}:
 \begin{equation}
 d_n \sim 4\times 10^{-4}\, \bigg(\frac{P}{1\, {\rm ms}}\bigg)^{1/2} \mbox{cm,} \quad
 d_p \sim 5\times 10^{-10}\, \bigg(\frac{\bar B}{10^{12}\,{\rm G}}\bigg)^{-1/2} \mbox{cm},
 \label{eq:disvor}
 \end{equation}
 respectively, where $P$ is the rotational period, and $\bar B$ is the average magnetic field.
 %The neutron intervortex distance is ruled by rotation
 %while proton intervortex distance is governed by the magnetic field.
 For example, for the Vela pulsar, whose measured rotational period is 89.33 
 ms, the spacing between neutron vortices is estimated to be 10$^{-3}$ cm, while, assuming a typical average magnetic field of 10$^{12}$ G, the distance between proton vortices
 should be around 10$^{-10}$ cm. This estimate indicates that proton-vortex (or fluxtube) arrays would densely fill up the space between neutron vortices. 
 %Along with the Fermi wavelengths and the healing lengths, the intervortex spacing give rise to different possible scenarios in the core of neutron stars. 
 %At microscopic (nuclear) scale in the outer core of neutron stars, 
 Then, it seems plausible to consider proton vortices lying between faraway neutron vortices.
 Similarly as in other multicomponent superfluids \cite{Minoru2011}, we will refer to this topological defect, a proton-superconductor vortex without overlapping neutron-superfluid vortex, as a half-quantized vortex in the two-component superfluid of the outer core. 
 In this case, the Fermi wavelength and the 
 healing distance in the superconducting fluid are typically of the order of 10 fm, much smaller than the intervortex proton spacing, which, in turn, is assumed to be several orders of 
 magnitude smaller than the neutron vortex separation. As a 
 consequence of the entrainment as well as of the nuclear interactions, the proton vortex is expected to produce a long wavelength perturbation in the much denser neutron superfluid.

 In the present paper, we contribute to the characterization of the structure and the effect of half-quantized vortices in the
two-component superfluid of the outer core of neutron stars within a generalized hydrodynamic description. The model includes the dynamical entrainment and the effective nucleon-nucleon interaction (of Skyrme type SLy4) between superfluid
neutrons and superconductor protons;
the dynamical variables are the density and velocity, gathered in a complex order parameter of the neutron and proton Cooper pairs.
Essentially the same model was recently used in the exploration of the long wave length excitations of the outer-core superfluid \cite{Gil2021}.
Since this approach cannot provide independent microscopic information on the differences between the density of fermions and the corresponding order parameter, we restrict our focus on study cases where the typical fermionic and bosonic length scales of the proton fluid approximately match. This fact allows us to consider proton-superconductor vortices characterized by a single length scale, which produce hydrodynamic perturbations on the underlying neutron superfluid.
The influence of realistic angular rotation and electromagnetic fields on the vortex structure is considered.

The paper is structured as follows: In Section \ref{sec:model}, first we derive the equations of motion in terms of the system's order parameters in the absence of rotation and magnetic field; these quantities are introduced later in the model to better show their effects.  In Section \ref{sec:results} we discuss the predictions of our model, present our numerical calculations of a proton vortex in the outer core, and compare with the literature. Finally, Section \ref{sec:final} presents the summary and prospects of future work. The Appendices \ref{sec:appendixA},  \ref{sec:appendixB},  \ref{sec:appendixC} , and  \ref{sec:appendixD} are devoted to featuring our choice for the effective nuclear interaction,  the effect of entrainment in the energy density through a toy model, the corresponding hydrodynamic equations of our system, and details of the performed numerical simulations, respectively.

\section{Model of the superfluid outer core} 
\label{sec:model}
Neutron and proton superfluids will be described by two
complex order parameters $\psi_n$ and $\psi_p$, respectively. From them, 
superfluid densities and velocity fields are obtained as $\rho_\sigma=|\psi_\sigma|^2$ , and $\vec{\rm v}_\sigma =\hbar \nabla\theta_\sigma/m$, respectively, where $\theta_\sigma=\arg \psi_\sigma$, and $m$ is the nucleon mass.
The system energy follows from the energy functional, $E[\psi_n,\psi_p]=\int \mathcal{H}\, d^3{r}$, whose energy density, in the absence of rotation and electromagnetic field, reads 
\begin{align}
\mathcal{H}_{pn}= \frac{1}{2\tilde m_n}|\hat p\,\psi_n|^2+\frac{1}{2\tilde m_p}|\hat p\,\psi_p|^2+\nu m\vec{j}_n\cdot\vec{j}_p+\epsilon_{nuc},
\label{eq:energy}
\end{align}
where $\tilde m_\sigma$ is a short notation for the effective masses $\tilde m_n=m/(1-\nu\rho_{p})$ and $\tilde m_p=m/(1-\nu\rho_{n})$ (see Appendix \ref{sec:appendixB}), $\hat p=-i\hbar\nabla$ is the canonical momentum operator, $\vec{j}_\sigma=\Re\{\psi_\sigma^*\hat p\psi_\sigma\}/m=\rho_\sigma \vec{\rm v}_\sigma$ are the current densities of each component when considered as separated fluids,  and $\nu$ (with units of volume) is the entrainment parameter that depends on the nuclear interaction, which is in turn encapsulated in the energy term $\epsilon_{nuc}=\epsilon_{nuc}(\rho_\sigma,\,\nabla\rho_\sigma)$ (see Appendix \ref{sec:appendixA} for details). 
Equation (\ref{eq:energy}) gathers interaction terms up to second order in the gradients of the order parameters \cite{Alpar1984}, and  explicitly features the contribution of entrainment to the kinetic energy, through the effective masses in the first two terms, and to the coupling between current densities, in the third term.
%Alternativelly, by gathering the entrainment terms, Eq. (\ref{eq:energy}) is rewritten as
%\begin{multline}
%\mathcal{H}=\frac{\hbar^2}{2m}\left[|\nabla\psi_n|^2+|\nabla\psi_p|^2\right]+\\ \nu\,\left[-\frac{\hbar^2}{2m}\left(\rho_p|\nabla\psi_n|^2+\rho_n|\nabla\psi_p|^2\right)
%+m\vec{j}_n\cdot\vec{j}_p \right]+\epsilon_{nuc},
%\label{eq:energy2}
%\end{multline}
%

Entrainment contributions are imposed by the Galilean invariance of the coupled system, which demands current densities $\vec J_\sigma$ for each component that involve both superfluid velocities $\vec{\rm v}_n$ and $\vec{\rm v}_p$ \cite{Andreev1976}, as
\begin{equation}
	\begin{aligned}
	\vec{J}_{n}= \vec{j}_{n}+ \vec j_{np} ,\\
	\vec{J}_{p}= \vec{j}_{p}-\vec j_{np},
	\end{aligned}
	\label{eq:current}
\end{equation}
where $\vec j_{np}=\nu\rho_n\rho_p(\vec{\rm v}_p-\vec{\rm v}_n)$ is the entrainment current density, and satisfy the continuity equations (see Appendix \ref{sec:appendixC})
\begin{equation}
\begin{aligned}
\frac{\partial\rho_n}{\partial t}+\nabla \vec J_n=0,\\
\frac{\partial\rho_p}{\partial t}+\nabla \vec J_p=0.
\end{aligned}
\label{eq:continuity}
\end{equation}
These expressions state, in this nonrelativistic approach, the independent conservation of the number of neutrons and protons. Notice that while the total particle current $\vec{J}=\vec{J}_{n}+\vec{J}_{p}=\vec{j}_{n}+\vec{j}_{p}=\rho_n \vec{\rm v}_n+\rho_p \vec{\rm v}_p$ does not depend on entrainment, the relative current density does:  $\vec{J}_{n}-\vec{J}_{p}=\vec{j}_{n}-\vec{j}_{p}+2 \vec j_{np}$.

The equations of motion follow from the variation of the total energy density Eq. (\ref{eq:energy}),  $\delta\mathcal{H}/\delta\psi_\sigma^*=i\hbar\partial_t\psi_\sigma$, which produces generalized, coupled Ginzburg--Landau (or Gross--Pitaevskii) equations
\begin{align}
i\hbar\frac{\partial\psi_n}{\partial t}=\mu_n\psi_n-\frac{\hbar^2}{2m}\nabla^2\psi_n \nonumber \\
-\frac{\hbar^2\nu}{2m}\left[|\nabla\psi_p|^2\psi_n-\nabla(\rho_p\nabla\psi_n)\right]-\frac{i\hbar\nu}{2}\left[\vec{j}_p\nabla\psi_n+\nabla(\vec{j}_p\psi_n)\right],
\label{eq:GPE1}
\\
i\hbar\frac{\partial\psi_p}{\partial t}=\mu_p\psi_p-\frac{\hbar^2}{2m}\nabla^2\psi_p  \nonumber \\
-\frac{\hbar^2\nu}{2m}\left[|\nabla\psi_n|^2\psi_p-\nabla(\rho_n\nabla\psi_p)\right]-\frac{i\hbar\nu}{2}\left[\vec{j}_n\nabla\psi_p+\nabla(\vec{j}_n\psi_p)\right],
\label{eq:GPE2}
\end{align}
where we have introduced the (generalized) chemical potentials
\begin{align}
\mu_\sigma=\frac{\partial\epsilon_{nuc}}{\partial\rho_\sigma}
-\nabla\frac{\partial\epsilon_{nuc}}{\partial\nabla\rho_\sigma},
\label{eq:mu}
\end{align}
 that include contributions from the density, $\partial\epsilon_{nuc}/\partial\rho_\sigma$, and the density-gradient, $\mathcal{Q}_\sigma=-\nabla(\partial\epsilon_{nuc}/\partial\nabla\rho_\sigma)$, dependence of the nuclear interaction energy SLy4; explicitly, the latter contributions are $\mathcal{Q}_p=-{\hbar^2}(\vartheta_{pp}^0 \nabla^2\rho_p + \vartheta_{np}^0 \nabla^2\rho_n)/(2m)$ and $\mathcal{Q}_n=-{\hbar^2}(\vartheta_{nn}^0 \nabla^2\rho_n + \vartheta_{np}^0 \nabla^2\rho_p)/(2m)$, where the coefficients $\vartheta_{ij}^0$ are obtained from the parametrization of the interaction (see Appendix \ref{sec:appendixA}). As a whole, the right hand sides of Eq. (\ref{eq:GPE1}-\ref{eq:GPE2}) reflect an involved coupling between the two superfluids due to entrainment effects.

\subsection{Rotating superfluids subjected to a magnetic field}
The star's rotation and magnetic field are introduced in the model, in a minimal coupling approach, through the kinetic (or mechanical) momentum operators
\begin{equation}
\begin{aligned}
\widehat{\Pi} = \hat p-m(\vec{\Omega}\times\vec{r}),\\
\widehat{\Pi}_A= \hat p-m(\vec{\Omega}\times\vec{r})-e\vec{A},
\end{aligned}
\label{eq:gen_momentum}
\end{equation}
where $\vec{\Omega}$ is the angular velocity of the star, $\vec{A}$ is the electromagnetic vector potential, and $e$ is the unit electric charge. By means of Eq. (\ref{eq:gen_momentum}), the energy density (\ref{eq:energy}) is just rewritten in the rotating frame of reference with the replacements $\hat p\psi_n\rightarrow \hat\Pi\psi_n$ and $\hat p\psi_p\rightarrow \hat \Pi_A\psi_p$, and
 the generalized currents of the independent superfluids, $\vec{j_n}=\Re(\psi_n^*\widehat{\Pi}\psi_n/m)=\rho_n (\hbar \nabla\theta_n/m-\vec{\Omega}\times\vec{r})=\rho_n \vec {\rm v}_n$ and 
$\vec{j_p}=\Re(\psi_p^*\widehat{\Pi}_A\psi_p/m)=\rho_p (\hbar \nabla\theta_p/m-\vec{\Omega}\times\vec{r}-e\vec A/m)=\rho_p \vec {\rm v}_p$. In addition (see Appendix \ref{sec:appendixB}), the energy density gets an energy shift given by 
\begin{align}
\Delta\mathcal{H}_{pn}=
-\psi_n^*\frac{m}{2}(\vec{\Omega}\times\vec{r})^2\psi_n-\psi_p^*\left[\frac{m}{2}(\vec{\Omega}\times\vec{r})^2+e\vec{A}(\vec{\Omega}\times\vec{r})\right]\psi_p.
\label{eq:Denergy}
\end{align}

On the other hand, to obtain the corresponding Maxwell equations, the energy of the electromagnetic field and the electron contributions have to be considered. Hence, the total energy density becomes
\begin{align}
\mathcal{H}=\mathcal{H}_{pn}+\Delta\mathcal{H}_{pn}+e(|\psi_p|^2-n_e)\,\Phi+e\vec J_e\cdot \vec A+\frac{|\vec B|^2}{2\muup_0},
\label{eq:Htotal}
\end{align}
where $\Phi$ is the scalar electromagnetic potential, $\muup_0$ is the magnetic permeability, and $(n_e,\,\vec J_e)$ are the electron density and electron current density, respectively.
The resulting generalized Ginzburg--Landau Eqs. (\ref{eq:GPE1}-\ref{eq:GPE2}), expressed in the rotating frame of reference, become
\begin{align}
i\hbar\frac{\partial\psi_n}{\partial t}=\mu_n\psi_n+\frac{\widehat{\Pi}^2\psi_n}{2m}-\frac{m}{2}(\vec{\Omega}\times\vec{r})^2\psi_n \nonumber\\
+\nu\left[\frac{|\widehat{\Pi}_A\psi_p|^2}{2m}\psi_n+\frac{1}{2}\vec{j_p}\widehat{\Pi}\psi_n+ 
\frac{1}{2}\widehat{\Pi}(\vec{j_p}\psi_n)+\frac{i\hbar}{2m}\nabla(\rho_p\widehat{\Pi}\psi_n)\right],
\label{eq:RotGPE1}
\\
i\hbar\frac{\partial\psi_p}{\partial t}=(\mu_p+e\Phi)\psi_p+\frac{\widehat{\Pi}_A^2\psi_p}{2m}-\left[\frac{m}{2}(\vec{\Omega}\times\vec{r})^2+e\vec{A}(\vec{\Omega}\times\vec{r})\right]\psi_p \nonumber\\
+\nu\left[\frac{|\widehat{\Pi}\psi_n|^2}{2m}\psi_p+ 
\frac{1}{2}\vec{j_n}\widehat{\Pi}_A\psi_p+ 
\frac{1}{2}\widehat{\Pi}_A(\vec{j_n}\psi_p)+\frac{i\hbar}{2m}\nabla(\rho_n\widehat{\Pi}_A\psi_p)\right].
\label{eq:RotGPE2}
\end{align}
while the resulting Ampere's law $\nabla \times \vec B =\muup_0 e\,(\vec J_p-\vec J_e)$ can be written in terms of the vector potential, in the Coulomb gauge $\nabla \cdot \vec A =0$, as 
\begin{align}
\nabla^2 \vec A = -\muup_0 e\,(\vec J_p-\vec J_e),
\label{eq:Ampere}
\end{align}
where $\vec J_e$ is the current density of electrons.
The coupled equations (\ref{eq:RotGPE1}-\ref{eq:Ampere}) have to be self-consistently solved for the set $\{\psi_n,\,\psi_p,\,\vec A\}$.

\subsection{Vortices}
 Vortices are featured by the quantized winding of the phase $\theta_\sigma$ around singular points that trace the vortex cores, where the phase is not uniquely defined. Since the order parameters are single-valued functions, the phase jumps in integer multiples of $2\pi$, that is $\oint \vec {d\ell}\cdot \nabla \theta_\sigma=2\pi q$, where $q=\pm 1,\pm2,\dots$ is the winding number. 
Usually, we will refer to singly quantized vortices, with $q=1$, because vortices with winding numbers larger than one are not energetically stable (see for instance Ref. \cite{Pitaevskii2003}).
 In the absence of rotation and magnetic field, the quantized winding of the phase around the vortex core translates also into a quantized velocity circulation $\Gamma_\sigma\equiv\oint \vec {d\ell}\cdot \vec {\rm v}_\sigma=(\hbar/m)\oint \vec {d\ell}\cdot \nabla \theta_\sigma=(2\pi\hbar/m)\,q$. However, this is not the general case when the kinetic momentum, as defined in Eqs. (\ref{eq:gen_momentum}), have additional contributions apart from the canonical momentum.

 {In the presence of rotation and magnetic field, energy minimization with respect to the velocity fields, assuming constant densities, leads to vanishing superfluid velocities $\vec{\rm v}_n=\vec{\rm v}_p=0$ \cite{Alpar1984}, hence to	}
\begin{align}
\frac{\hbar}{m} \nabla \theta_n=\vec{\Omega}\times\vec{r},
\label{eq:london1} \\
\frac{\hbar}{m} \nabla \theta_p-\frac{e}{m}\vec{A}=\vec{\Omega}\times\vec{r}.
\label{eq:london2}
\end{align}
 { Equation (\ref{eq:london1}) is complied with by means of quantum vortices in the neutron superfluid, so it generates vorticity through singular phase lines in the neutron phase $\theta_n$ that balance the otherwise unbalanced vorticity $2\vec \Omega$ introduced by the angular velocity. The ratio between the total flux of this vorticity (or equivalentely, the circulation of the velocity) and the circulation per vortex $\Gamma_n$ determines the inter-vortex distance $d_n$ of Eq. (\ref{eq:disvor}).}
	{  On the other hand, the gauge invariance of the proton wave function prevents the appearance of proton vortices due to rotation in the absence of an overall magnetic field (that would have $\vec A=0$). Instead, the generation of surface currents gives rise to the so-called London's magnetic field $\vec B_L$, which [from taking the curl of Eq. (\ref{eq:london2}) for an emergent vector potential $\vec A_L$] and by fulfilling}
\begin{equation}
\frac{e\vec{B_L}}{m} = -{2\vec{\Omega}},
\label{eq:londonfield}
\end{equation}
 {is capable of balancing rotation, so that the protons do not move in the rotating frame. Due to the typical rotation rates $|\Omega|\sim 100$ Hz, this magnetic field takes small values, $|B_L|\sim 0.1-1$ G, in comparison with the typical magnetic fields of neutron stars inferred from observations. The latter fields are believed to be trapped in the superconducting star interior since early, non-superconducting stages of the star \cite{Baym1969}, and have to satisfy Ampere's law Eq. (\ref{eq:Ampere}). }
  Because of entrainment, the current density $\vec J_p$ includes a contribution from the superfluid neutron velocity, which contributes also to the total charge current density and so to the overall stellar magnetic field.
	The associated magnetic flux threads the star outer core through flux tubes along the core of quantum vortices in the proton superconductor (assumed to behave as a type II superconductor \cite{Baym1969}). 
	The ratio between the total magnetic flux (as computed by the average magnetic field $\bar B$ inferred from observations) and the flux quantum per vortex $\phi_0=2\pi\hbar/e$ determines the proton inter-vortex distance $d_p$ in Eq. (\ref{eq:disvor}). The angular frequency of the cyclotron orbits, $\omega_B$, provides the characteristic time $\omega_B^{-1}=(e|\bar B|/m)^{-1}$ and length $\ell_B=\sqrt{\hbar/(m\omega_B)}=\sqrt{\hbar/(e|\bar B|)}$ scales of the magnetic field. 
	
\begin{figure}[tb]
	\includegraphics[width=1\linewidth]{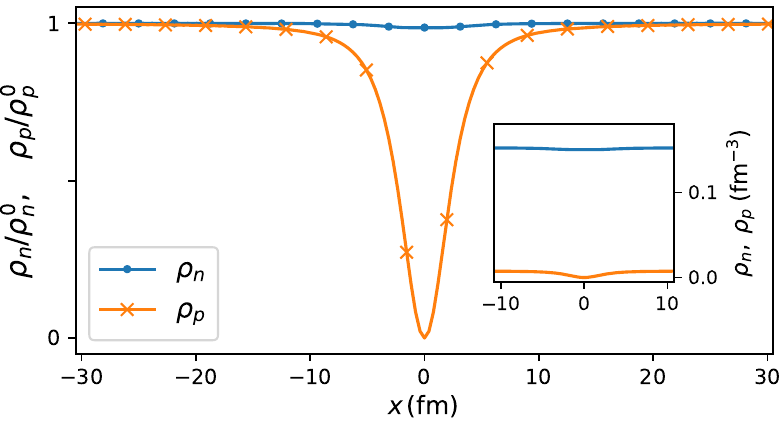}
	\caption{Cross section of a half vortex state in the superfluid-superconductor mixture of neutrons and protons at typical nuclear-matter density in the outer core of neutron stars. The vortex is excited in the proton superconductor and produces a hydrodynamic perturbation in the much denser neutron superfluid. The main graph shows scaled densities with respect to the respective bulk-density values of neutron and protons, $\rho_n^0$ and $\rho_p^0$, whereas the inset shows the density profiles in absolute units.}
	\label{fig:halfvortex}
\end{figure}

  Proton supercurrents around vortex lines preserve superconductivity in the presence of strong magnetic fields, which enter the bulk only up to small length scales $\Lambda$. This can be seen, by assuming constant densities and point-like vortices in Eq. (\ref{eq:Ampere}), and using Eqs. (\ref{eq:current}) and (\ref{eq:gen_momentum}), in the resulting London equation for the magnetic field (see e.g. Ref. \cite{Tinkham})
\begin{equation}
\nabla ^2 \vec{B} -\frac{\vec{B}}{\Lambda ^2} = -\frac{\phi_0}{\Lambda ^2}\delta(\vec r)-\muup_0{e}\,\nu\rho_n\rho_p 2\vec{\Omega},
\label{eq:penetration_length}
\end{equation}
 with a source term that consists of a proton vortex and a neutron vorticity (assumed to take the average value $2\vec \Omega$).
	The London penetration depth is calculated as  
$\Lambda=\sqrt{{m}\,[{\muup_0 e^2\rho_p(1-\nu\rho_n)}]^{-1}}$, and takes values of the order $\Lambda\sim$ 74 fm at $\rho=0.16$ fm$^{-3}$ (or $\sim$50 fm at higher density $\rho=0.24$ fm$^{-3} $\cite{Alpar1984}). As a result, magnetic-field maxima of the order of $B_v \sim \phi_0/(2\pi\Lambda^2)\sim 10^{15}$ G $\gg \bar B$ are estimated to peak at proton-vortex cores, as expected from a penetration length ($\Lambda\propto \sqrt{\bar B/B_v}\, d_p\sim 10^{-2}\,d_p$) significantly smaller than the distance between proton vortices $d_p$. 

 \begin{figure}[tb]
	\includegraphics[width=1\linewidth]{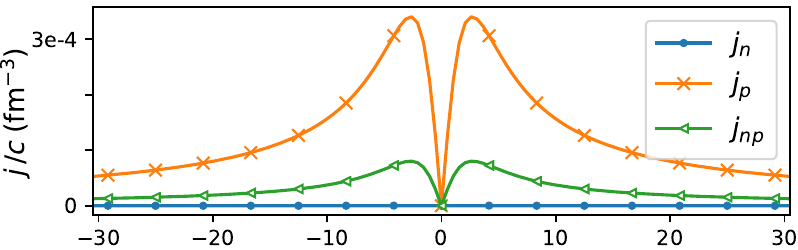}
	\includegraphics[width=1\linewidth]{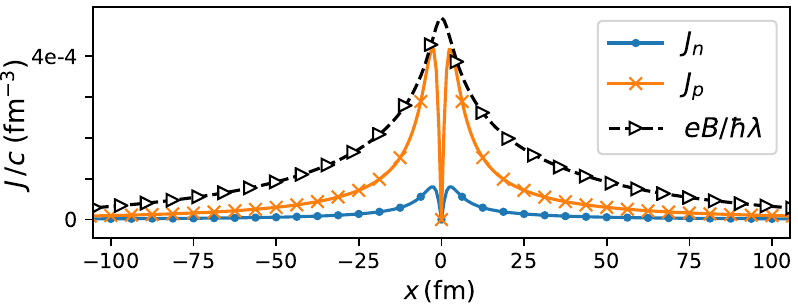}
	\caption{Particle current densities corresponding to the proton vortex shown in Fig. \ref{fig:halfvortex}. The top panel depicts the currents of the independent neutron $\vec j_n$ and proton $\vec j_p$ fluids, along with the entrainment current $\vec j_{np}$. The botton panel shows the current densities of neutron $\vec J_n$ and protons $\vec J_p$ in the coupled system, as defined in Eqs. (\ref{eq:current}-\ref{eq:continuity}), along with the magnetic $\vec B$ field threading the vortex [where $\lambda=2\pi \hbar/(mc)$ is the Compton wavelength]. }
	\label{fig:halfvortexJ}
\end{figure}
\section{Proton vortices in the outer core}
\label{sec:results}

Although neutron and proton vortices can overlap producing a quantum vortex that wind the total phase of the system twice ($2\pi$ per component) around the vortex core, single vortex excitations that winds only one of the two component phases (so just $2\pi$ in one component's phase and no winding in the other) are also possible. This latter configuration is the object of our study, and is referred to as half quantized vortex or just half vortex. From a symmetry-based  perspective, the difference in phase winding is associated with the existence of two $U(1)$ symmetries, so an overall $U(1)\otimes U(1)$ symmetry, which are broken by the Bose condensed fluids of neutrons and protons (see for instance Ref. \cite{Minoru2011}).

We focus on a study case where the typical fermionic, $\lambda_{Fp}$, and bosonic, $\xi_p$, length scales of the proton superconductor are the same (as obtained within the Skyrme SLy4 interaction). 
As can be seen in Fig. \ref{fig:lengths}, such a matching occurs deep in the outer core for a nucleon density of $\rho\sim 0.16$ fm$^{-3}$, where  $\lambda_{Fp}\approx\xi_p = 10.3$ fm. Due to the strong asymmetry, $\delta=0.9$, the presence of a proton vortex is expected to produce just a hydrodynamic perturbation in the underlying neutron superfluid, whose typical length scale is much smaller $\lambda_{Fn}= 3.8$ fm. Under these conditions, the structure of a proton vortex can be plausibly characterized within our model, and so we will analyze straight vortex lines that are assumed to be invariant along the $z-$coordinate (which we ignore from now on).

\begin{figure}[tb]
	\includegraphics[width=0.9\linewidth]{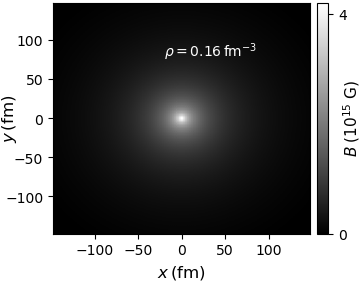}
	\caption{Single magnetic field flux tube threading the proton vortex described in Figs. \ref{fig:halfvortex} and \ref{fig:halfvortexJ}.}
	\label{fig:fluxline}
\end{figure}
A proton vortex centered at position $\vec r_0=(x_0,y_0)$ in an $XY$ plane,  can be described by an order parameter $\psi_p(\vec r)=\sqrt{\rho_p(\vec r)}\,\exp(i\varphi)$, whose phase is given by the angle around the center $\theta_p=\varphi=\arctan[ { (y- y_0)}/{(x-x_0)}]$, whereas its density vanishes at the vortex core  $\rho_p(\vec r_0)=0$, and, far from it,  for $|\vec r-\vec r_0|/\xi_p\gg 1$, recovers the bulk proton density $\rho_p^0=7.66\times 10^{-3}$ fm$^{-3}$. With these constraints,  Fig. \ref{fig:halfvortex} shows our numerical results after solving Eqs. (\ref{eq:RotGPE1}-\ref{eq:Ampere}) for a proton vortex subjected to angular rotation rate $\Omega=100$ Hz and average magnetic field $|\bar B|=10^{12}$ G (see Appendix \ref{sec:appendixD} for details about the numerical method).  The neutron superfluid, as expected, shows only a very small density depletion ($\sim 1 \%$ with respect to the bulk value $\rho_n^0=0.152$ fm$^{-3}$) at the position of the proton vortex. 
The observed depletion is mainly due to the the entrainment coupling between components.
Since close to the half-vortex axis the entrainment contributes with a nonzero current density to the stationary neutron continuity equation \eqref{eq:continuity}, this extra current is accompanied by a corresponding neutron density reduction in order for the equation to be satisfied.
With regard to the stellar rotation and magnetism, despite their apparent high values, their influence on the vortex structure is negligible. We have checked that the vortex structure shown in Fig. \ref{fig:halfvortex} remains unaltered for varying rotation rates and magnetic fields within ranges of realistic values  $10^2-10^3$ Hz and $10^{12}-10^{15}$ G, respectively. This results (or absence of effects) can be understood by computing the respective typical length scales, $\ell_\Omega=\sqrt{\hbar/(m|\Omega|)}=2.5\times 10^{10}$ fm and $\ell_B=2.6\times 10^3$ fm, which reflect negligible energy contributions (proportional to $\ell^{-2}$) against the typical vortex energies (proportional to $\xi_p^{-2}$).

 Characteristic values of the fluid velocity and current density around the vortex core can be estimated from ${\rm v}_p\sim\hbar/(m\xi_p)=0.02$ c,  $j_p=\rho_p {\rm v}_p\sim 10^{-4}$ fm$^{-3}$c, and  $j_{np}\sim\nu\rho_n\rho_p {\rm v}_p\sim -0.24\, j_p$, so that the overall proton current density estimate is $J_p\sim 10^{-4}$ fm$^{-3}$c; the magnetic field has nearly no contribution to the velocity, since  ${\rm v}_B= \ell_B \omega_B\sim 8\times 10^{-5}$ c. Our numerical results produce current peak values that are consistent with these estimates and are accompanied by long tails, as can be seen in  Fig. \ref{fig:halfvortexJ}, which shows the current density and magnetic field profiles around the proton vortex depicted in Fig. \ref{fig:halfvortex}. The bottom panel represents the total neutron and proton current densities, whose contributions [according to Eq. (\ref{eq:current})] are separately plotted in the top panel. The long tails show the peak currents reduced to half their values at distances $\xi_p\sim 10$ fm away from them, and are consistent with slower decays of the magnetic field according to the larger penetration length  $\Lambda\sim 70$ fm  estimated after Eq. (\ref{eq:penetration_length}). The magnetic flux produced by this varying magnetic field gives rise to one quantum of flux $\phi_0$ per vortex, the same flux associated with the observed average field. As can be seen, due to the entrainment,  non-negligible neutron currents settle around the proton vortex, which by means of the continuity equation are translated into the small neutron-density variations observed in Fig. \ref{fig:halfvortex}.
 
 Although our results have been obtained for the typical core density of $\rho=0.16$ fm$^{-3}$, some general features can be extrapolated to other densities in the core whenever the proton vortex size (proton coherence length) is clearly larger than the neutron fermi wave length. In these situations, the neutron density is expected to experience similar small depletions due to the entrainment effects produced by the proton vortex. On the contrary, when proton and neutron vortices overlap, both fermionic densities are expected to become significantly depleted due to the overall high current density affecting the whole system.
It is also worth commenting on the differences between a proton vortex and a neutron vortex (as estimated in Ref. \cite{Alpar1984}), since in the latter the threading magnetic flux is reduced by a factor of $\nu\rho_n/(1-\nu\rho_n)$, thus by $0.19$ at $\rho=0.16$ fm$^{-3}$,  with respect to the former \cite{Alpar1984}. Therefore, while we obtain magnetic-field peaks of magnitude $B_v \approx 4\times 10^{15}$ G  at the proton-vortex core, lower values would be expected at the corresponding neutron-vortex core. Figure \ref{fig:fluxline} depicts the carpet plot of the magnetic field showing the constrained flux around the vortex core located at the origin $\vec r_0=(0,0)$.

\section{Conclusions}
\label{sec:final}

Since typical rotation rates of neutron stars give rise to a low density of vortices in the outer-core neutron superfluid, most of the flux tubes produced by the star magnetic field in the accompanying proton superconductor thread single proton vortices that do not have a neutron vortex partner. We have analyzed the structure of these half vortices in a typical configuration of a type-II superconductor, where the London penetration length $\Lambda$ is much larger than the superconductor healing length $\xi_p$. By means of a generalized hydrodynamic model that includes the effects of neutron-proton entrainment and Skyrme SLy4 interactions, and through numerical simulations that search for stationary quantum vortex states, we have characterized the microscopic structure of the half vortex in a region of the star's outer core with nuclear density $\rho=0.16$ fm$^{-3}$, where the typical length scales of proton pairing and Fermi wave number match, $k_{Fp}= \xi_p\approx 10$ fm. It is shown how an isolated proton vortex produces a perturbative depletion in the underlying neutron superfluid, and generates particle currents that are increased by the nondissipative dragging of neutrons. Although the exponential decay of the magnetic field away from the vortex core extends up to typical lengths of several London penetration lengths, $\Lambda\approx 70$ fm, over which one quantum of magnetic flux $\phi_0$ and typical average values of $\bar{B}=10^{12}$ G are achieved, the length of the magnetic field's tail is still orders of magnitude smaller than the expected distance between neutron vortices $d_n\sim 10^{-4}$ cm. Despite the apparent very high  values (although consistent with observations) of both the rotation and the magnetic field considered, and due to the (even higher) nuclear densities of the outer core, the revealed structure of the vortex remains practically unaltered with respect to varying (or even the absence of) both rotation and magnetic field intensity. 

Our generalized hydrodynamic model provides a mesoscopic picture of the interplay between superfluid neutrons and superconductor protons in the outer core. As such, this picture does not include damping effects that can be directly associated with macroscopic nonequilibrium phenomena observed in the dynamics of neutron stars, like glitches.
Nevertheless, there is an incresing interest to discern the cause of this latter phenomenon, assuming that it is connected with neutron vortices pinned to the inner crust, and there is no clear role of the star's outer core \cite{Chamel2012,Andersson2012,Watanabe2017}. In this direction, the present work contributes to understand features of the outer core that can be relevant for its overall dynamics. In addition, , our model can contribute to characterize aspects of the thermal history of neutron stars similarly as in Ref. \cite{Kobyakov2017disp}.
Prospects of future work involve this problem along with further microscopic characterization of single half vortices by means of Hartree-Fock-Bogoliubov models, where both the fermionic and the bosonic sectors of the coupled Fermi systems are included, so that fermion densities and pairing fields can be separately revealed.

\appendix
 \section{SLy4 interaction}
 \label{sec:appendixA}
 
These Skyrme forces describe effectively the nuclear interaction
and provide the equation of state of a neutron star \cite{Vautherin1972,Bender2003}. Its energy density
\begin{align}
	\epsilon_{nuc}=\mathcal{H}^{\rho}+\mathcal{H}^{\nabla},
	\label{eq:Hskyrme}
\end{align}
contains terms that depend on the nuclear densities $\mathcal{H}^{\rho}(\rho_\sigma)$ and also terms that depend on the gradients of the neutron and proton densities $\mathcal{H}^{\nabla}(\nabla\rho_\sigma)$, and so the latter vanish for uniform density distributions but simulate otherwise finite-range effects of the interaction. In turn,
\begin{align}
	\mathcal{H}^{\rho}=\mathcal{T}+\mathcal{H}_{0}+\mathcal{H}_{3}+\mathcal{H}_{eff},
	\label{eq:Hrho}
\end{align}
where $\mathcal{T}$ is a kinetic term,
$\mathcal{H}_{0}$ is a zero-range two-body term,
$\mathcal{H}_{3}$ is a three-body term, and
$\mathcal{H}_{eff}$ is an effective-mass term.
The pairing contribution to the total energy of the system is not included assuming that it is negligible
against the typical Fermi levels.
More explicitly, 
\begin{align}
	\mathcal{T}=\dfrac{\hbar^{2}}{2m}(\tau_{n}+\tau_{p}),
\end{align}
accounts for the kinetic energy 
densities $\tau_{i}={3}\,(3\pi^{2})^{2/3}\rho_{i}^{5/3}/5$ of the Fermi gas,
whereas the other terms in Eq. (\ref{eq:Hskyrme}) provide the bulk part of the potential energy
\begin{align}
\mathcal{H}_{0}=\dfrac{t_{0}}{4}\left[(2+x_{0})\rho^{2}-(2x_{0}+1)(\rho_{p}^{2}
+\rho_{n}^{2})\right],
\end{align}
\begin{align}
\mathcal{H}_{3}=\dfrac{t_{3}\,\rho^{{1}/{6}}}{24}\left[(2+x_{3})\rho^{2}-(2x_{3}
+1)(\rho_{p}^{2}+\rho_{n}^{2})\right],
\end{align}
\begin{align}
\mathcal{H}_{eff}&=\left[t_{1}(2+x_{1})+t_{2}(2+x_{2})\right]
\dfrac{\rho(\tau_{n}+\tau_{p})}{8} \nonumber\\&
+\left[t_{2}(2x_{2}+1)-t_{1}(2x_{1}+1)\right]\dfrac{\tau_{p}\rho_{p}+\tau_{n}
\rho_{n}}{8},
\end{align}
and, as anticipated, the last term in Eq. (\ref{eq:Hskyrme}) is due to the short range of the nucleon-nucleon interaction,
\begin{align}
	\mathcal{H}^{\nabla}&= \frac{3}{32} [t_{1}(1-x_{1})-t_{2}(1+x_{2})] \,
	\left[(\nabla \rho_n)^2 + (\nabla \rho_p)^2\right]\nonumber\\&
	+ \frac{1}{16} [3t_{1}(2+x_{1})-t_{2}(2+x_{2})]\,\nabla \rho_n \cdot \nabla \rho_p.
\end{align}
From this latter term, to compute the contributions to the generalized chemical potential \eqref{eq:mu}, we define $\vartheta_{nn}^0=\vartheta_{pp}^0={3m}\, [t_{1}(1-x_{1})-t_{2}(1+x_{2})]/({8\hbar^2})$ and $\vartheta_{np}^0=\vartheta_{pn}^0={m}\,  [3t_{1}(2+x_{1})-t_{2}(2+x_{2})]/({8\hbar^2})$.

The parameters in these expressions are \cite{Chabanat1998,Douchin2001}:
$t_{0}=-$2488.91 MeV fm$^{3}$, $t_{1}=$486.82 MeV fm$^{5}$, $t_{2}=-$546.39 
MeV fm$^{5}$, $t_{3}=$13777.0 MeV fm$^{4}$, $x_{0}=$0.834, $x_{1}=-$0.344, 
$x_{2}=-$1.0, and $x_{3}=$1.354. From them,
following  Ref. \cite{Chamel2006}, the entrainment parameter $\nu$ can also be calculated and takes the value $\nu \approx -1.566$ fm$^{3}$.

The $\beta$ equilibrium in the bulk \cite{Baym1971} can be locally determined 
in terms of the asymmetry $\deltaup=({\rho_{n}-\rho_{p}})/{\rho}$ by means of Eq. (\ref{eq:Hskyrme}) and  the condition $\mu_n=\mu_p+\mu_e$, with $\mu_e=\hbar\,c\,(3\pi^2 \rho_e)^{1/3}$ for ultrarelativistic electrons and assuming constant densities and local electric-charge equilibrium $\rho_e=\rho_p$, as
\begin{align}
\left(\dfrac{\hbar^{2}}{2m}+ \frac{C_1}{8}\rho\right)\left(3\pi^{2} \dfrac{\rho}{2}\right)^{2/3} 
\left[(1+\deltaup)^{2/3}-(1-\deltaup)^{2/3}\right] \nonumber\\
-\dfrac{t_{0}}{2}\rho\,\deltaup\,(1+2x_{0}) 
-\dfrac{t_{3}}{12}(2x_{3}+1)\deltaup\,\rho^{7/6} \nonumber\\
+\dfrac{3\pi^{2}}{5}C_2\left(\dfrac{\rho}{2}\right)^{5/3}
\left[(1+\deltaup)^{5/3}-(1-\deltaup)^{5/3}\right] \nonumber\\
-\hbar c\left[3\pi^2\frac{\rho}{2}\,(1-\deltaup)\right]^{1/3}=0 ,
\label{eq:beta}
\end{align}
where $C_1=t_{1}(2+x_{1})+t_{2}(2+x_{2})$ and $C_2=t_{2}(2x_{2}+1))-t_{1}(2x_{1}+1)$. 
By solving this equation the density asymmetry $\deltaup$, that is the amount of neutrons and protons in the bulk, is obtained for a given total nuclear density $\rho$.

 \section{Classical analogue of entrainment in the presence of rotation and magnetic field}
\label{sec:appendixB}
To illustrate the transformation of the energy density Eq. (\ref{eq:energy}) in the presence of rotation and magnetic field, we consider a toy, classical model of two point particles with coupled velocities whose dynamics is described by the Lagrangian
\begin{align}
\mathcal{L}[r_1,r_2,\dot r_1,\dot r_2]=\frac{m_{11}\dot r_1^2}{2} + \frac{m_{22}\dot r_2^2}{2}+{m_{12}\dot r_1\dot r_2}+eA\dot r_2,
	\label{eq:toy_Lagrangian}
\end{align}
where $A$ is the vector potential of a magnetic field that couples only to one of the particles, and $m_1=m_{11}+m_{12}$, $m_2=m_{22}+m_{12}$; the condition $\Delta_m=m_1m_2-m_{12}^2>0$ is assumed.
If the whole system, its center of mass, is brought into a steady motion with velocity $V_0$, the Lagrangian can be rewritten in the reference frame moving with $V_0$ as $\mathcal{L}'[r_1',r_2',\dot r_1',\dot r_2']$, with new velocities $\dot r'_1=\dot r_1- V_0$ and $\dot r'_2=\dot r_2- V_0$. The corresponding momenta are
\begin{align}
p'_1=\frac{\partial \mathcal{L}'}{\partial \dot r'_1 }=m_{11}\dot r'_1+m_{12}\dot r'_2+m_1 V_0,\\
p'_2=\frac{\partial \mathcal{L}'}{\partial \dot r'_2 }=m_{22}\dot r'_2+m_{12}\dot r'_1+m_2 V_0+eA,
\label{eq:toy_momentum}
\end{align}
and then the Hamiltonian is obtained from $H'[r'_1,r'_2,p'_1,p'_2]=\sum p_\sigma\dot r_\sigma-\mathcal{L}'$ to be
\begin{align}
H'=\frac{{p'_1}^{2}}{2\tilde m_{11}} +\frac{(p'_2-eA)^2}{2\tilde m_{22}}+\frac{p'_1\,(p'_2-eA)}{\tilde m_{12}}\nonumber\\
-\alpha V_0(p'_1+p'_2)+(\alpha-1)\left(eA V_0+\frac{m_1+m_2}{2}V_0^2\right),
\label{eq:toyH}
\end{align}
where new effective masses arise $\tilde m_{ij}=\Delta_m/m_{ij}$, with $i,j=1,2$, and the parameter
 $\alpha=[m_1m_2-m_{12}(m_1+m_2)]/\Delta_m$ determines how the Hamiltonian transforms.
If $m_{12}=0$ then $\alpha=1$, and one recovers the usual transformation for the Hamiltonian in the moving frame $H\rightarrow H-V_0 P$, with total momentum $P=p_1+p_2$. A non-vanishing coupling mass $m_{12}$ produces the more involved transformation shown in Eq. (\ref{eq:toyH}), which can be rewritten as
\begin{align}
H'=\frac{{\Pi'}_1^2}{2\tilde m_{11}} +\frac{{\Pi'}_2^2}{2\tilde m_{22}}+\frac{{\Pi'}_1\,{\Pi'}_2}{\tilde m_{12}}- eA\, V_0-\frac{m_1+m_2}{2}V_0^2,
\label{eq:toyH1}
\end{align}
in terms of the kinetic momenta
\begin{align}
\Pi'_1=p'_1-m_1\,V_0,\\
\Pi'_2=p'_2-m_2\,V_0-eA.
\label{eq:toyPi}
\end{align}

 \section{Hydrodynamic equations}
  \label{sec:appendixC}
  \begin{figure}[tb]
  	\includegraphics[width=1\linewidth]{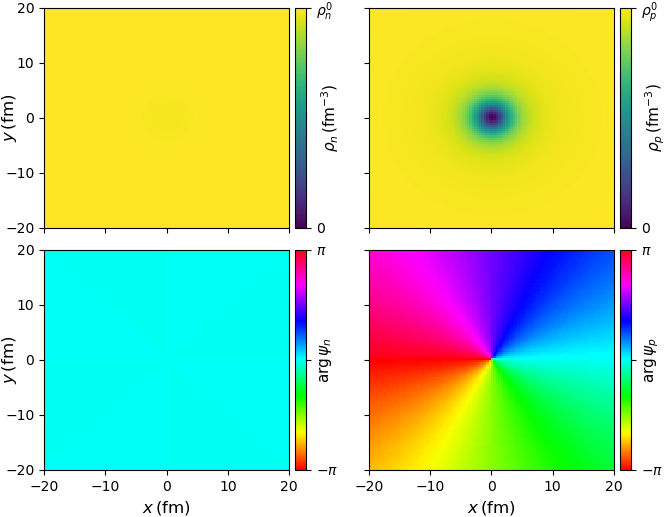}
  	\caption{Number density (top panels) and phase (bottom panels) profiles of the neutron (left) and proton (right) order parameters around the proton-vortex core.}
  	\label{fig:phases}
  \end{figure}

 The hydrodynamic equations can be derived directly from the generalized time-dependent Ginzburg--Landau Eqs. (\ref{eq:RotGPE1}-\ref{eq:RotGPE2}) by writing the order parameters in the so-called Madelung transformation as $\psi_\sigma=\sqrt{\rho_\sigma(\vec{r},t)}\exp{i\theta_\sigma(\vec{r},t)}$ and defining the superfluid velocities of the decoupled superfluids as $\vec v_\sigma=\hbar \nabla \theta_\sigma/m$. From the separation of the real and imaginary parts of the resulting expressions, one obtains equations of motion for superfluid densities and phases in the rotating frame. On the one hand, from the imaginary terms, one recovers the continuity Eqs. (\ref{eq:continuity}) that can be rewritten as
\begin{equation}
\begin{aligned}
\frac{\partial\rho_n}{\partial t}+\nabla\left[
\rho_n\vec {\rm v}_n+\nu\rho_n\rho_p(\vec {\rm v}_p-\vec {\rm v}_n)
\right]=0,
\\
\frac{\partial\rho_p}{\partial t}+\nabla\left[
\rho_p\vec {\rm v}_p-\nu\rho_n\rho_p(\vec {\rm v}_p-\vec {\rm v}_n)
\right]=0;
\end{aligned}
\label{eq:continuity2}
\end{equation}
on the other hand, after taking the gradient of the resulting real terms, one gets the momentum equations
\begin{widetext}
\begin{align}
\frac{\partial}{\partial t}\left(\vec{\rm v}_n+\vec{\Omega}\times\vec{r}\right)+\nabla \left\lbrace \frac{\mu_n}{m} -\frac{\hbar^2}{m^2}\left[\frac{(1-\nu\rho_p)\nabla^2\sqrt{\rho_n}}{2\sqrt{\rho_n}}+\frac{\nu}{2}(\nabla\sqrt{\rho_p})^2-\frac{\nu\nabla\rho_p\nabla\rho_n}{4\rho_n}\right] \hspace{5cm}\nonumber \right.\\\left.
+\frac{1}{2}\left[\vec{\rm v}_n^2-(\vec{\Omega}\times\vec{r})^2-\nu\rho_p(\vec{\rm v}_p-\vec{\rm v}_n)^2-\nu\rho_p\vec{\rm v}_p(\vec{\Omega}\times\vec{r})\right]  \right\rbrace=0  , \hspace{2cm}
\label{eq:phase_n}
\\
\frac{\partial}{\partial t}\left(\vec{\rm v}_p+\vec{\Omega}\times\vec{r}+\frac{e\vec{A}}{m}\right)+\nabla\left\lbrace \frac{\mu_p+e\Phi}{m}
-\frac{\hbar^2}{m^2}\left[\frac{(1-\nu\rho_n)\nabla^2\sqrt{\rho_p}}{2\sqrt{\rho_p}}+\frac{\nu}{2}(\nabla\sqrt{\rho_n})^2-\frac{\nu\nabla\rho_p\nabla\rho_n}{4\rho_p}\right] \hspace{4cm}\nonumber \right.\\\left.
+\frac{1}{2}\left[\vec{\rm v}_p^2-(\vec{\Omega}\times\vec{r})^2-\frac{e \vec{A}}{m}(\vec{\Omega}\times\vec{r})-\nu\rho_n(\vec{\rm v}_p-\vec{\rm v}_n)^2-\nu\rho_n\vec{\rm v}_n\left(\vec{\Omega}\times\vec{r}+\frac{e\vec{A}}{m}\right)
\right] \right\rbrace =0 .\hspace{1cm}
\label{eq:phase_p}
\end{align}
\end{widetext}
While the continuity Eqs. (\ref{eq:continuity2})  are related to the conservation of the number of particles in each superfluid component, the phase Eqs. (\ref{eq:phase_n}--\ref{eq:phase_p}) are related to the corresponding conservation of momentum.

 \section{Numerical simulations}
\label{sec:appendixD}

The stationary vortex states have been calculated within a cylindrical computational domain, with the vortex core aligned with the cylinder axis, and the cylinder walls determined by imposed potentials $W(\vec r)=10\mu_n\,\{1+\tanh[5k_{Fn}(|\vec r|-R)]\}$ in Eqs. (\ref{eq:RotGPE1}. \ref{eq:RotGPE2}). The cylinder radius $R$ has been fixed by the condition of reaching both the bulk density values, $\rho_n^0$ and $\rho_p^0$, of neutron and proton superfluids in $\beta-$equilibrium, and non-varying magnetic fields faraway from the vortex core. Our results have been tested by achieving convergence in the densities, currents, and magnetic field values against variations of the cylinder radius and the number of computational grid points $N_x\times N_y$; typical values of these parameters were $R\in [100, 500]$ fm and $N_x\times N_y\in [512\times 512,\, 1024\times 1024]$ points.
In the search for stationary states, the initial state was fixed by the bulk nuclear density values and by imprinting a vortex phase in the proton order parameter, that is,
$\psi_n(x,y)=\max[ 1-W(x,y)/\mu_n, 0 ]\times \sqrt{\rho_n^0}$ and $\psi_p(x,y)=\max[ 1-W(x,y)/\mu_p, 0 ]\times \sqrt{\rho_p^0}\,\exp[i\arctan(y/x)]$; a vanishing density was also set at the vortex core $\psi_p(0,0)=0$. Starting with this state, an imaginary time evolution of Eqs. (\ref{eq:RotGPE1},\ref{eq:RotGPE2}), self consistently with Eq. \eqref{eq:Ampere}, was performed until convergence by means of standard finite-difference spatial discretization and time integrators based in Runge-Kutta methods coded in Julia programming language. An example of the density and phases of the converged states is shown in Fig. \ref{fig:phases}.

\bibliography{Vortex_paper_v4}
\bibliographystyle{apsrev4-1}

\end{document}